\documentclass[%
 reprint,
 amsmath,amssymb,
 aps,
pra,superscriptaddress
]{revtex4-1}

\usepackage{bm}
\usepackage{graphicx}
\usepackage[usenames]{color}
\usepackage{physics}
\usepackage{siunitx}
\usepackage{xcolor}
\usepackage{setspace}
\usepackage{float}
\usepackage{tabularx}
\usepackage[linkcolor=blue,citecolor=blue,colorlinks=true,urlcolor=blue]{hyperref}

\newcommand{\be}[1]{\begin{equation}\label{#1}}
\newcommand{\ee}{\end{equation}}
\usepackage{amsmath,esint}
\usepackage{comment}
\usepackage{lipsum}
\usepackage[toc]{appendix}
\usepackage[T1]{fontenc}
\usepackage{soul}

\begin{document}

\preprint{APS/123-QED}

\title{Quantum versus semi-classical signatures of correlated triple ionization in Dalitz plots}
\author{Dmitry K. Efimov}
\affiliation{Institute of Theoretical Physics, Faculty of Fundamental Problems of Technology,
Wrocław University of Science and Technology, 50-370 Wrocław, Poland}
\author{Georgios P. Katsoulis}
\affiliation{Department of Physics and Astronomy, University College London, Gower Street, London WC1E 6BT,
England, United Kingdom}
\author{Tymoteusz Rozp\k{e}tkowski}
\author{Sergiusz Chwa\l{}owski}
\affiliation{Institute of Theoretical Physics, Faculty of Fundamental Problems of Technology,
Wrocław University of Science and Technology, 50-370 Wrocław, Poland}
\author{Agapi Emmanouilidou}
\affiliation{Department of Physics and Astronomy, University College London, Gower Street, London WC1E 6BT,
England, United Kingdom}
\author{Jakub S. Prauzner-Bechcicki}
\email{jakub.prauzner-bechcicki@uj.edu.pl}
\affiliation{Jagiellonian University in Kraków, Faculty of Physics, Astronomy and Applied Computer Science,
Marian Smoluchowski Institute of Physics, Łojasiewicza 11, 30-348 Krakow, Poland}

\date{\today}
\begin{abstract}
We investigate correlated three-electron escape in Ne when driven by an intense, infrared laser field. We do so by employing a reduced-dimensionality quantum-mechanical model 
and two three-dimensional semi-classical models.  One semi-classical model is a recently developed one that accounts with effective coulomb potentials for the  interaction between two bound electrons (ECBB) while it fully accounts for all other interactions. The other semi-classical model  is the Heisenberg one, which effectively accounts for the interaction of each electron with the core via a  soft-core potential. We identify and compare the signatures of correlated three-electron escape for both quantum and semi-classical models on Dalitz plots and find a better agreement between the quantum and the ECBB model. We also show that a central ``spot'' on the Dalitz plots is reproduced by all models. Using  the ECBB model we associate this ``spot'' with the direct triple ionization pathway and argue this to be the case  also for the quantum model. Devising a simple classical 
model that accounts for the direct pathway of triple ionization, we show that the width of this spot in the Dalitz plots solely depends on the time of tunnel-ionization. 
\end{abstract}

\maketitle

\section{\label{sec:intro}Introduction}
Studies of interaction of strong laser fields with atoms and molecules brought out many intriguing phenomena: starting with above-threshold ionization, non-sequential multiple ionization and high-harmonic generation, through attosecond pulse generation, to initiate eventually a spectrum of new experimental techniques, just to mention attosecond streaking based measurements or attosecond transient spectroscopy~\cite{Borrego-Varillas2022}.
Progress in experimental works has always been accompanied by a considerable effort aimed at the theoretical description of studied phenomena.
Achieving the latter is still quite a challenge, especially when in the focus are processes where more than two electrons are involved.
Non-sequential multiple ionization (NSMI) is an excellent example.
While description of non-sequential double ionization may be considered settled~\cite{Corkum93,Walker94,Watson97,Lein00,parker2000time,sacha2001pathways,Parker06,staudte2007binary,rudenko2007correlated,Faria11,becker2012theories,Bergues12,Katsoulis2018}, the description of triple ionization cannot be considered as such, regardless of the effort that has been put into addressing it~\cite{Sacha2001,Ruiz05,ho2006plane,ho2007argon,Guo08,Thiede2018,Prauzner-Bechcicki2021,Efimov2021,Jiang2022}. 
There are several complementary approaches to the problem. One may try to use strong field approximation, however, already for two electrons the calculation becomes very complicated, not mentioning inclusion of a third electron~\citep{Amini2019}.
Next, one may use classical and semi-classical methods~~\cite{Sacha2001,ho2006plane,ho2007argon,Katsoulis2018,PhysRevA.105.043102} or directly tackle the numerical solution of an appropriate Schr\"odinger equation~\cite{parker1998intense,Lein00,parker2000time,Parker06,ruiz2006ab,Prauzner-Bechcicki2008,Thiede2018}.
In both, classical and quantum, approaches one either uses the full geometry of the problem or uses simplified models.
Yet, in each case the singularity of the Coulomb potential poses a problem.

Out of the multitude of possible strong-field phenomena, we want to focus our attention on NSMI. In brief, NSMI is considered a three-step process ~\cite{Corkum93}: (i) an electron tunnel-ionizes from the laser-field-lowered Coulomb potential, (ii) this electron is accelerated in the laser field and can return to rescatter from the parent ion and (iii) transfers energy to the bound electrons leading to multiple ionization. The first step is not allowed classically, therefore it is typically included via the quantum-mechanical Ammosov-Delone-Krainov formula~\cite{Landau,Delone:91} leading to a semi-classical approach. In modelling the second and the third steps, semi-classical methods are founded on solving the Hamilton's or Newton's equations of motion. In principle, for semi-classical methods, the number of active electrons is not as challenging as for quantum approaches, yet the Coulomb singularity in the electron-core interaction term is problematic. The latter allows an electron to acquire a negative energy of any value, which may be compensated via electron-electron interaction through artificial ionization. Therefore, one introduces the soft-core Coulomb-like potentials~\cite{Haan1994} or Heisenberg potentials~\cite{heisen1,heisen2} as a remedy. Let us clarify, the Heisenberg potential (H model) effectively leads to softening of the Coulomb singularity~\cite{PhysRevA.105.043102,PhysRevA.107.L041101}. However, softening the Coulomb potential in classical models does not accurately describe electron scattering from the core \cite{Pandit2018,Pandit2017}. One way to address this is to fully account for the Coulomb interaction of each electron with the core as well as for the Coulomb interaction between any pair of electrons that are not both bound. In this case, effective Coulomb potentials are used to account for the interaction of a bound-bound electron pair (ECBB model) \cite{PhysRevA.105.043102}.

The quantum mechanical approach includes all three steps of the process by default, however, increasing the number of electrons involved becomes a formidable task, especially when the Schr\"odinger equation is solved on a gird. Therefore, full-dimensional quantum calculations on a grid have been performed only for two electrons and for a specific range of parameters of the laser pulse (i.e. frequency, field amplitude, phase, envelope)~\cite{parker1998intense,parker2000time,Parker06,feist2008nonsequential,Hao14}. Models incorporating reduced geometries are used to allow greater flexibility concerning the parameters of the laser pulses and the number of electrons involved~\cite{Lein00,ruiz2006ab,Prauzner-Bechcicki2008,Efimov2018,Thiede2018}. In these models, typically, each of the electrons is allowed to move along a one-dimensional track and  jointly with the use of the soft-core Coulomb-like potential make the problem computationally tractable. Using the value of the   soft-core parameter  to adjust the ionization energy is an additional benefit of such an approach.

In the following we shall use both, semi-classical and quantum, approaches to describe triple ionization of atoms in strong laser fields. Our goal is to study ``fingerprints'' of correlated electron escape found in the final momenta  of the escaping electrons. Collecting experimental data for triple and higher ionization is still a difficult task when it comes to measuring momenta of all electrons and the nucleus; it may be feasible in the future~\cite{Basnayake2022,Grundmann20,Larimian20,Mikaelsson20,Zhong20,Henrichs18,Bergues12}. To visualize the momenta of the three outgoing electrons and identify the electron-electron correlations, we map all relevant events onto Dalitz plots~\cite{Efimov2021,Jiang2022}. We find that a middle spot in the Dalitz plots of the electron momenta is a fingerprint of correlated  electron-escape which is reproduced in a consistent manner regardless of the approach taken, i.e. semi-classical or quantum. 

The paper is structured as follows, first, we briefly introduce the semi-classical and quantum models employed in this work. Then, we discuss the obtained electron momenta distributions in the context of Dalitz plots and compare the semi-classical with the quantum results. Next, we examine signatures of direct ionization, i.e. simultaneous emission of all three electrons following recoliision, in the Dalitz plots.  We identify a spot in the middle of Dalitz plots as a signature of direct ionization. Very importantly, we devise a simple classical model to explain the width of this spot and connect this width with the time the recolliding electron tunnel-ionizes through the field-lowered Coulomb potential.

\section{Description of the models}

\subsection{Classical models}
We employ two three-dimensional (3D) semiclassical models of NSMI developed in the non-dipole framework, accounting for the motion of the core and the three electrons. In what follows, we describe the formulation of the ECBB model and the H model that address multi-electron escape in strongly driven atoms. The two methods resolve in a different way unphysical autoionization in 3D semiclassical models that fully account for the Coulomb singularity, with the ECBB model having better agreement with experiment \cite{PhysRevA.107.L041101}.

\subsubsection{Effective Coulomb potential method}
 The cornerstone of the Effective Coulomb potential model \cite{PhysRevA.105.043102,PhysRevA.107.L041101} is the exact treatment of the two interactions that we consider the most important during a recollision. Namely, we account for the full Coulomb potential between each electron,  bound or quasifree, and the core. Quasifree refers to a recolliding electron or an electron escaping to the continuum. We also treat the exact Coulomb potential, and hence the exact energy transfer between any pair of a quasifree and a bound electron.

To tackle the problem of artificial autoionization, we take a different approach to softening the Coulomb potential. Specifically, we use effective Coulomb potentials to account for the interaction of a bound-bound electron pair (ECBB), i.e., we approximate the energy transfer from a bound to another bound  electron. Hence, we expect the ECBB model to be more accurate for laser pulse parameters where multi-electron ionization due to transfer of energy between electrons in excited states after recollisions plays less of a role. A sophisticated element of the ECBB model is its ability to classify an electron as quasifree or bound on the fly during time propagation. That is, we decide on the fly if the interaction between a pair of electrons will be described by the full or an effective Coulomb potential. To do so,  we use a set of simple criteria detailed below.

The Hamiltonian of the three-electron atom is given by
\begin{equation}\label{Hamiltonian_effective}
\begin{aligned}
&H = \sum_{i=1}^{4}\frac{\left[\mathbf{\tilde{p}}_{i}- Q_i\mathbf{A}(y,t) \right]^2}{2m_i}+\sum_{i=2}^{4}\frac{Q_iQ_1}{|\mathbf{r}_1-\mathbf{r}_i|} \\
&+\sum_{i=2}^{3}\sum_{j=i+1}^{4} \left[ 1-c_{i,j}(t)\right]\frac{Q_iQ_j}{|\mathbf{r}_i-\mathbf{r}_j|} +\sum_{i=2}^{3}\sum_{j=i+1}^{4}c_{i,j}(t) \\
&\times\Big[V_{\mathrm{eff}}(\zeta_j(t),|\mathbf{r}_{1}-\mathbf{r}_{i}|) + V_{\mathrm{eff}}(\zeta_i(t),|\mathbf{r}_{1}-\mathbf{r}_{j}|)\Big],
\end{aligned}
\end{equation}
where $Q_i$ is the charge, $m_i$ is the mass, $\mathbf{r}_{i}$ is the position vector and $\mathbf{\tilde{p}}_{i}$ is the canonical momentum vector of particle $i$. The mechanical momentum $\mathbf{p}_{i}$ is given by
\begin{equation}
\mathbf{p}_{i}=\mathbf{\tilde{p}}_{i}-Q_i\mathbf{A}(y,t).
\end{equation}
where $\mathbf{A}(y,t)$ is the vector potential and $\mathbf{E}(y,t) = - \dfrac{\partial\mathbf{A}(y,t)}{\partial t}$ is the electric field. The effective Coulomb potential that an electron $i$ experiences at a distance $|\mathbf{r}_{1}-\mathbf{r}_{i}|$ from the core (particle 1 with $Q_{1}=3$), due to the charge distribution of electron $j$ is derived as follows \cite{PhysRevA.40.6223,PhysRevA.105.043102}. We approximate the wavefunction of a bound electron $j$ with a 1s hydrogenic wavefunction
\begin{equation}
\psi(\zeta_j,|\mathbf{r}_{1}-\mathbf{r}_{j}|) = \left( \frac{\zeta_j^3}{\pi} \right)^{1/2} e^{-\zeta_j |\mathbf{r}_{1}-\mathbf{r}_{j}|},
\end{equation}
with $\zeta_j$ the effective charge of particle $j$ \cite{PhysRevA.105.043102,PhysRevA.40.6223}. Hence, using Gauss's law, one finds that the potential produced due to the charge distribution $-|\psi(\zeta_j,|\mathbf{r}_{1}-\mathbf{r}_{j}|) |^2$ is given by
\begin{equation}
V_{\mathrm{eff}}(\zeta_j,|\mathbf{r}_{1}-\mathbf{r}_{i}|) =  \dfrac{1-(1+\zeta_j| \mathbf{r}_{1}-\mathbf{r}_{i}|)e^{-2\zeta_j| \mathbf{r}_{1}-\mathbf{r}_{i}|}}{| \mathbf{r}_{1}-\mathbf{r}_{i}|},
\end{equation}
with $\zeta_j$ the effective charge of particle $j$ \cite{PhysRevA.40.6223,PhysRevA.105.043102}. When $ \mathbf{r}_{i}\rightarrow\mathbf{r}_{1}$, the effective potential is equal to $\zeta_j$, hence ensuring a finite energy transfer between bound electrons $i$ and $j$. As a result no artificial autoionization takes place. The functions $c_{i,j}(t)$ determine at time $t$ during propagation whether  the full Coulomb or  effective $V_{\mathrm{eff}}(\zeta_i,|\mathbf{r}_{1}-\mathbf{r}_{j}|)$ and $V_{\mathrm{eff}}(\zeta_j,|\mathbf{r}_{1}-\mathbf{r}_{i}|)$ potentials describe the interaction between electrons $i$ and $j$  \cite{PhysRevA.105.043102}. The effective potentials are activated only when both electrons in a pair are bound. The simple criteria we use to determine if an electron is bound or quasifree is as follows. A quasifree electron can  transition to bound following a recollision. Specifically, after a quasifree electron  has its closest approach to the core, it is considered bound if its position along the $z$ axis is influenced more by the core than the electric field. On the other hand, a bound electron can transition to quasifree due to transfer of energy during a recollision or from the laser field.  In the former case, a bound electron becomes quasifree if its potential energy with the core constantly decreases following recollision. A bound electron can also transition to quasifree due to the laser field if its energy  becomes and remains positive. The criteria are discussed in detail and illustrated in Ref. \cite{PhysRevA.105.043102}.

We use a vector potential of the form
\begin{equation}\label{eq:vector_potential}
\mathbf{A}(y,t) = -\frac{E_0}{\omega}\exp \left[ - 2\ln (2)\left( \frac{c t - y}{c \tau} \right)^2 \right]   \sin ( \omega t  - k y) \hat{\mathbf{z}},
\end{equation}
where $k=\omega/c$ is the  wave number of the laser field.  The direction of the vector potential and the electric field is along the $z$ axis, while the direction of light propagation is along the $y$ axis.  The magnetic field, $ \mathbf{B}(y,t) = \grad \times \mathbf{A}(y,t)$, points  along the $x$ axis.  The pulse duration is $\tau = 25$ fs, while  the wavelength is  800 nm. For Ne, we consider intensities 1.0, 1.3 and $\mathrm{1.6 \; PW/cm^2 }$. The  highest intensity  considered here, is chosen such that  the probability for a second electron to tunnel ionize solely due to the laser field is very small \cite{PhysRevA.107.L041101}. Hence,  electron-electron correlation prevails in triple ionization, with  the bound electrons ionizing only  due to recollisions.

One electron tunnel ionizes through the field-lowered Coulomb barrier at time $t_0$ along the direction of the total laser field. Tunneling occurs with a non-relativistic quantum-mechanical tunneling rate described by the instantaneous Ammosov-Delone-Krainov (ADK) formula  \cite{Landau,Delone:91} with the empirical corrections by Tong and Lin for high intensities \cite{Tong_2005}. Using this formula, we obtain a rate that also accounts for depletion of the initial ground state \cite{PhysRevA.107.L041101}. We find the time $t_0$ in the time interval [-2$\tau$, 2$\tau$] where the electric field is nonzero, using importance sampling \cite{ROTA1986123}, with $\tau$  the full width at half maximum of the pulse duration in intensity. The exit point of the recolliding electron along the direction of the electric field is obtained analytically using parabolic coordinates \cite{HUP1997533}. The electron momentum along the electric field is set equal to zero, while the transverse one is given by a Gaussian distribution. This distribution represents the Gaussian-shaped filter with an intensity-dependent width arising from standard tunneling theory \cite{Delone:91,Delone_1998,PhysRevLett.112.213001}. For the initially bound electrons, we employ a micro-canonical distribution \cite{PhysRevA.105.043102}, while the core is initially at rest.

In our formulation, we fully account for the Coulomb singularities. Hence, an electron can approach infinitely close to the nucleus during time propagation. To ensure the accurate numerical treatment of the N-body problem in the laser field, we perform a global regularisation. This regularization was introduced in the context of the gravitational N-body problem  \cite{Heggie1974}. To integrate Hamilton's equations of motion, we use a leapfrog technique  \citep{Pihajoki2015,Liu2016} jointly with the Bulirsch-Stoer method \cite{press2007numerical,bulirsch1966numerical}. This leapfrog technique allows integration of Hamilton's equation when the  derivatives of the positions and the momenta  depend  on the quantities themselves. The steps involved in this technique, employed in this work, are described in detail in Ref. \cite{PhysRevA.105.043102}.

\subsubsection{Heisenberg potential method}
An alternative approach for excluding unphysical autoionization in 3D semiclassical treatments is adding a Heisenberg potential for the interaction of each electron with the nucleus \cite{PhysRevA.21.834}.  This amounts to adding a potential barrier that mimics the Heisenberg uncertainty principle and prevents each electron from a close encounter with the nucleus.  The advantage of this model is that it describes electronic interactions  via Coulomb forces at all times during time propagation. However, the reduction of the phase space that each electron can access does not accurately describe the interaction of each electron with the nucleus, leading to ``softer" re-collisions upon the return of the recolliding  electron to the core \cite{PhysRevA.105.043102}.

The Heisenberg potential, originally proposed by Kirschbaum and Wilets in Ref. \cite{PhysRevA.21.834}, is given by
\begin{equation}\label{Heisenberg}
V_{H,i}=\dfrac{\xi^2}{4\alpha \mu r_{i,1}^2}\exp \left\{ \alpha \left[ 1 - \left(  \dfrac{ r_{i,1} p_{i,1} }{\xi}  \right)^4 \right]  \right\},
\end{equation}
where $ \mathbf{r}_{i,1}=\mathbf{r}_1 - \mathbf{r}_i  $ is the relative position of each one of the three electrons $i$ = 2, 3, 4 with respect to the core $i$ = 1, $\mathbf{p}_{i,1}$ is the corresponding relative momentum
\begin{equation}
\mathbf{p}_{i,1} = \dfrac{m_i\mathbf{p}_1 - m_1\mathbf{p}_i}{m_1 + m_i},
\end{equation}
 and $\mu = m_1m_i/(m_i+m_1) $ is the reduced mass of the electron core system.
This potential restricts the relative position and momentum of electron $i$ according to
\begin{equation}\label{Constraint}
r_{i,1} p_{i,1}  \geq \xi.
\end{equation}
Hence, the Heisenberg potential acts as a repulsive potential when the electron is close to the nucleus.

Including the Heisenberg potential for all electron-core pairs, we find that the Hamiltonian is given by
\begin{equation}\label{Hamiltonian_heis}
\begin{aligned}
H = \sum_{i=1}^{4}\frac{\left[\mathbf{\tilde{p}}_{i}- Q_i\mathbf{A}(y,t) \right]^2}{2m_i} + \sum_{i=1}^{3}\sum_{j=i+1}^{4}\frac{Q_i Q_j}{|\mathbf{r}_i-\mathbf{r}_j|}  + \sum_{i=2}^{4}V_{H,i}.
\end{aligned}
\end{equation}

\subsection{Quantum model}

The numerical treatment of three-electron atoms in a complete configuration space poses a significant challenge. We employ a model with restricted dimensionality capable of capturing the core dynamics of electrons in a strong field \cite{Sacha2001,Thiede2018}. In this model, each electron moves along a one-dimensional (1D) track inclined at a constant angle $\alpha$ with respect to the laser pulse's polarization axis ($\tan \alpha= \sqrt{2/3}$). The tracks form a constant angle of $\pi/6$ between each pair. This geometry is determined through a local stability analysis of the adiabatic full-dimensional potential \cite{Sacha2001,eckhardt06,arnold2013mathematical} and has been successfully applied in previous studies of three-electron dynamics \cite{Thiede2018,Efimov2019,Efimov2020,Prauzner-Bechcicki2021,Efimov2021,Efimov2023}.

The Hamiltonian is given by:
\begin{equation}
	\label{hamiltonian}
	H = \sum_{i=1}^3 \frac{p^2_i}{2} + V + V_{int},
\end{equation}
where $V$ and $V_{int}$ represent the atomic and interaction potentials, respectively. The atomic potential in the restricted space is expressed as:
\begin{equation}
	V = -\sum_{i=1}^3 \frac{3}{\sqrt{r^2_i+\epsilon^2}}+\sum_{i,j=1;i<j}^3\frac{q^2_{ee}}{\sqrt{(r_i-r_j)^2+r_ir_j+\epsilon^2}}
\end{equation}
with a smoothing factor $\epsilon=\sqrt{0.83}$ and an effective charge $q_{ee}=1$ to reproduce the triple ionization potential of the Neon atom ($I_p=4.63$ a.u.). The interaction term is described as:
\begin{equation}
	V_{int} = \sqrt{\frac{2}{3}} A(t)\sum_{i=1}^3 p_i,
\end{equation}
where the vector potential is defined as

$$A(t)=\frac{F_0}{\omega}\sin^2\left(\frac{\pi t}{T_p}\right)\sin\left(\omega t + \phi\right)$$

\noindent for $t\in [0,T_p]$. The laser pulse parameters are: field amplitude, $F_0$; carrier frequency, $\omega = 0.06$ a.u.; pulse length $T_p=2\pi n_c/\omega$; number of optical cycles, $n_c=3$; and carrier-envelope phase, $\phi$, set to zero.

To study the interaction of three-electron atoms with a laser pulse, we need a wavefunction describing the motion of three electrons (assuming the nucleus is infinitely heavy). The wavefunction is composed of spatial and spin parts and  requires antisymmetry with respect to electron exchange. For a simplified Hamiltonian described in eq.~(\ref{hamiltonian}) and an atom with $s^2p^1$, two electrons have the same spin, providing the spin part possesses symmetry in the exchange of those electrons. Thus, the spatial part must be antisymmetric in that exchange.

Without loss of generality, the wavefunction for such an atom in the restricted-space model can be written~\cite{Efimov2023}:
\begin{multline}\label{wf}
	\Psi\propto \Psi_{12}(r_1,r_2,r_3)|{\rm UUD}\rangle \\
	+\Psi_{23}(r_1,r_2,r_3)|{\rm DUU}\rangle\\
	+\Psi_{13}(r_1,r_2,r_3)|{\rm UDU}\rangle.
\end{multline}
Here, $U$ and $D$ represent electrons with spin-up and spin-down, respectively. The subscripts denote the electron pair with respect to exchange of which the wavefunction is antisymmetric.

The Hamiltonian, eq.(\ref{hamiltonian}), does not affect the spin part of the wavefunction during evolution. Thus, for simplicity in numerical implementation, it suffices to evolve only one of the three terms on the right side of eq.(\ref{wf}). During calculations, we evolve the wavefunction $\Psi_{12}(r_1,r_2,r_3)$, which is antisymmetric with respect to the exchange of electrons $1$ and $2$, but neither symmetric nor antisymmetric with respect to the exchange of electrons $1\leftrightarrow 3$ and $2\leftrightarrow 3$. The initial wavefunction, representing the ground state, is obtained using imaginary time propagation of the Hamiltonian, eq.(\ref{hamiltonian}), without the interaction with the field and enforcing the proper symmetry~\cite{Thiede2018}.

The numerical solution of the time-dependent Schr\"odinger equation (TDSE) with the Hamiltonian, eq.(\ref{hamiltonian}), extends the method used in the two-dimensional case described elsewhere~\cite{Prauzner-Bechcicki2008,Efimov2021}. The TDSE is solved on a large 3-dimensional grid using a split-operator technique and the fast Fourier transform algorithm. For obtaining momentum distributions of outgoing electrons, the wavefunction cannot be absorbed at the grid's edges. Thus, we divide the evolution space into ``bounded motion'' and ``outer'' regions. The evolution in the ``bounded motion'' region proceeds without further simplifications, while in the ``outer'' regions, the Hamiltonian is successively simplified by neglecting the interaction of electrons with the nucleus and other electrons. This allows representation and evolution of the wavefunction in the momentum representation in the ``outer'' regions, simplifying the evolution to the multiplication by an appropriate phase factor. The transfer between the ``bounded motion'' and ``outer'' regions is achieved through smooth cutting and coherent adding of the wavefunction, following the procedure introduced in Ref~\cite{Lein00}.

At the simulation's end, the wavefunction from ``outer'' regions can be integrated over the ``bounded'' part, yielding momentum distributions corresponding to single ionization, double ionization, and triple ionization.
The grid has $n=1024$ nodes in each direction, with a step size of ${\rm d}r=0.195$. The time-step is ${\rm d}t=0.05$, and the total number of steps amounts to $6500$.
Further details of the algorithm for simulating momentum distributions of the three-electron atom can be found in the supplementary material of Ref \cite{Efimov2021}.

\subsection{Dalitz Plots}
\begin{figure}[t] 
	 \includegraphics[width=0.4\textwidth]{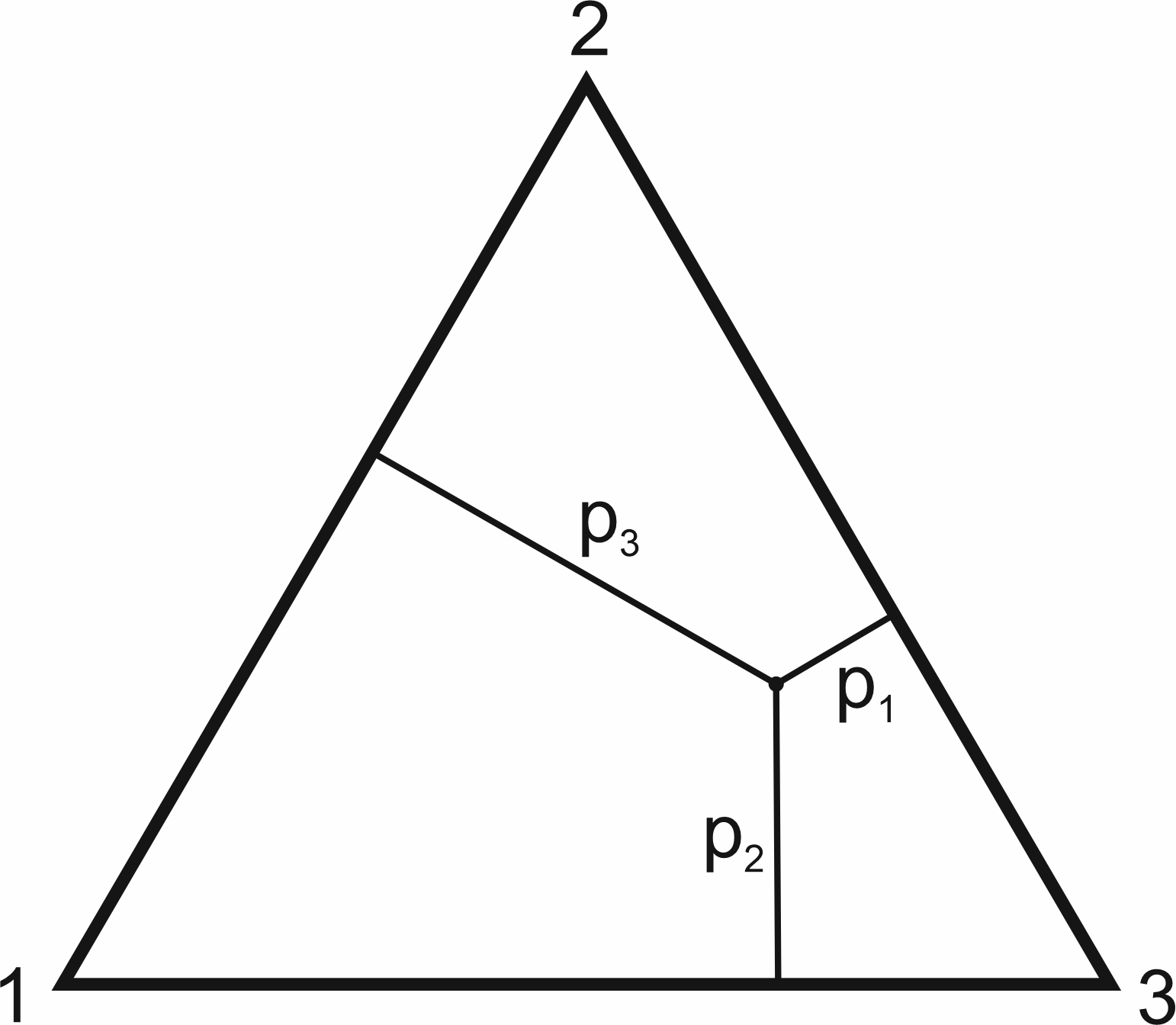}
	 \caption{Schematic visualization of a point in the ternary plot.\label{ternary}}
\end{figure}

To visualize three-electron distributions we use the so-called Dalitz or ternary plots~\cite{Efimov2021,Jiang2022}. These are obtained by first projecting the data onto the sphere of radius $R=\sqrt{p_1^2+p_2^2+p_3^2}$, where $p_i$ is the $i$-th electron's momentum along the polarization axis. Then, using gnomonic projection onto planes tangent to the sphere the final ternary plots are reached in each octant of the momentum space. Ternary plots are read in the following way (see Fig.~\ref{ternary}): the vertices of the triangle are denoted $1, 2, 3$ and correspond to momentum components along the polarization axis of three electrons, i.e. the point of the vertex is equivalent with the statement that the given electron carried all momentum. Positions of the points inside the triangle are defined by a set of three distances $p_1, p_2, p_3$ to the sides of the triangle opposite to vertices $1,2,3$, respectively. The closer to the side the point is the smaller momentum is carried by the given electron. In the momentum space spanned by momenta components along the polarization axis there are two types of octants available, those in which all electrons are oriented in the same direction and those in which one of the electrons has opposite orientation with respect to other two. In the first category there are two octants, marked as (+++) or (-- -- --). In the second category there are six octants, marked as (++ --) and all possible permutations thereof.

\subsection{Direct vs. Delayed}
In the classical models, we register a triple ionization (TI) event as direct if three electrons ionize shortly after recollision. We register a triple ionization event as delayed if one electron ionizes with a delay after recollision. For TI,
we identify direct  and delayed events as follows \cite{PhysRevA.105.043102,PhysRevA.107.L041101}:
\begin{enumerate}
\item{We find the ionization time of each of the three electrons, $t^i_{\text{ion}}$.}
\item{ We register the maxima in the inter-electronic potential energies as a function of time between electron pairs $i,\; j$ and $i,\; k$ and $j,\; k$ during the time intervals when in these pairs one electron is quasifree and the other is bound. Next, for each electron $i$, we identify the maximum for each one of the $i,\; j$ and $i,\; k$ potential energies that is closest to the time $t^i_{\text{ion}}.$ We denote these times as $t_{\text{rec}}^{i,j}$ and $t_{\text{rec}}^{i,k}$. We obtain at most six such times for TI events.}
\item{For each time $t_{\text{rec}}^{i,j}$ we identify the time $t_2$ of closest approach to the core of the quasifree electron (either electron $i$ or $j$)  that is closest to $t_{\text{rec}}^{i,j}$ and denote it as $t^{i,j}_2$. We obtain at most six such times for TI events.}
\end{enumerate}
We label an event as direct or delayed TI if four of the times $t^{i,j}_2$ are the same, accounting for one electron being quasifree and the other two bound. That is, if electron $i$ is quasifree during the recollision closest to the ionisation time $t^i_{\text{ion}}$ then the times $t^{i,j}_2$, $t^{i,k}_2$, $t^{j,i}_2$ and $t^{k,i}_2$ should be the same. The times $t^{j,i}_2$ and $t^{k,i}_2$ are associated with the recollision times $t^{j,i}_{\text{rec}}$ and $t^{k,i}_{\text{rec}}$ for the bound electrons $j$ and $k$, respectively. For the quasifree electron we obtain two recollision times $t^{i,j}_{\text{rec}}$ and $t^{i,k}_{\text{rec}}$ associated with the ionization time $t^i_{\text{ion}}$. We choose  the one with the largest difference from  $t^i_{\text{ion}}$, guaranteeing a stricter criterion for direct TI events.
 Next, we label a TI event as direct  if the following  conditions are satisfied $\Delta t_{1}=|t^{i,j}_{\text{rec}} - t^i_{\text{ion}}| < t_{\text{diff}}$ or $(t^i_{\text{ion}} < t^{i,j}_{\text{rec}} \; \text{and} \; t^i_{\text{ion}} < t^{i,k}_{\text{rec}})$ and $\Delta t_{2}=|t^{j,i}_{\text{rec}} - t^j_{\text{ion}}| < t_{\text{diff}}$ and $\Delta t_{3}=|t^{k,i}_{\text{rec}} - t^k_{\text{ion}}| < t_{\text{diff}}$. The condition $(t^i_{\text{ion}} < t^{i,j}_{\text{rec}} \; \text{and} \; t^i_{\text{ion}} < t^{i,k}_{\text{rec}})$ has also been used in our previous studies \cite{ChenA2017Ndiw,Slingshot} to account for a quasifree electron ionising significantly earlier before recollision. This happens mostly at high intensities.   We label events as delayed pathway TI, when two electrons ionize shortly after recollision, while one electron ionizes with a delay. That is, for these delayed events  one out of the three times $\Delta t_{1}$, $\Delta t_{2}$ and $\Delta t_{3}$ is larger than $t_{\text{diff}}$ and the other two  times are less than $t_{\text{diff}}$.
The time  $t_{\text{diff}}$ is determined by the time interval where the interelectronic potential energy undergoes a sharp change due to a recollision. For the intensities considered here, we find  $t_{\text{diff}} \approx T/8$, with $T$ being the period of the laser field.

\section{Results and discussions}
\subsection{General Observations}
\begin{figure}[t]
	 \includegraphics[width=0.5\textwidth]{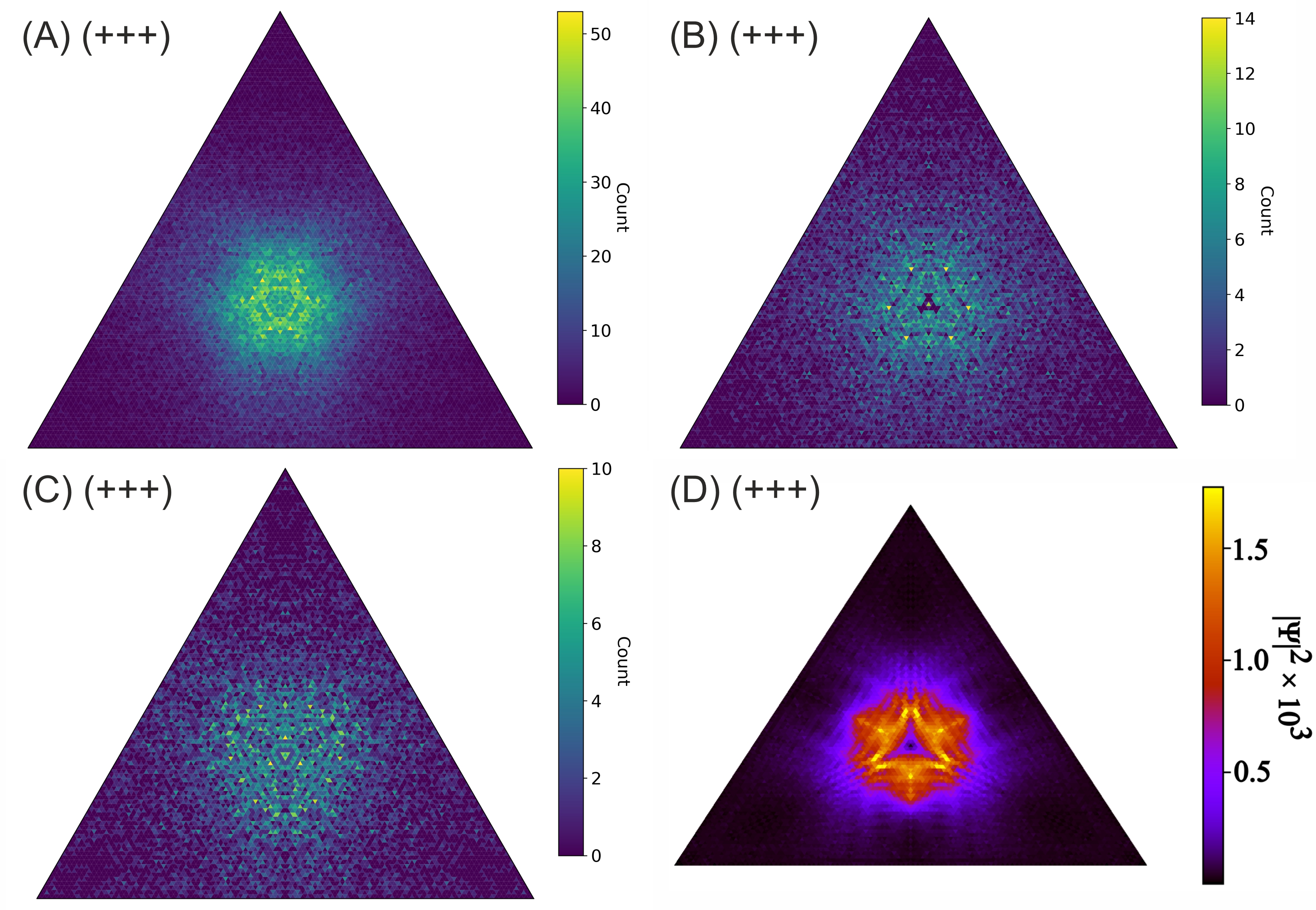}
	 \caption{Momenta distributions for triple ionization events visualized with the help of Dalitz plots (ternary plots): (A) semi-classical calculations with the ECBB model, $I= 1.6$ ${\rm PW/cm}^2$, (B) semi-classical calculations with the H model, $\alpha=2$, $I= 1.6$ ${\rm PW/cm}^2$, (C) semi-classical calculations with the H model, $\alpha=4$, $I= 1.6$ ${\rm PW/cm}^2$, (D) quantum calculations, $I= 1.7$ ${\rm PW/cm}^2$.\label{main_feature}}
\end{figure}
Let us begin our discussion by presenting the main features of correlated triple ionization on Dalitz plots, see Fig.~\ref{main_feature}. Panels (A)-(C) correspond to results obtained with the two semi-classical models and panel (D) to results obtained with the quantum model. For the distribution presented in panel (A) we used the ECBB model, whereas for panels (B) and (C) we used the H model with different values of the parameter $\alpha$; for the quantum case the soft-core 1D+1D+1D potential is used. We find that for each of the plots in Fig.~\ref{main_feature} the momentum distribution is concentrated and peaked in the middle of the triangle, and stretches towards its sides. 

The interpretation of the observed distribution is rather straightforward. The peak in the middle of the triangle corresponds to dominance, for the chosen laser intensity, of events in which all three electrons escape in the same hemisphere, i.e. the electrons are escaping in the same direction along the polarization of the laser field, with similar momenta. The tails of the distribution that stretch towards the sides of the triangle correspond to events in which one of the three electrons has significantly smaller momentum  compared to the other two electrons, which escape with similar momenta.
\begin{figure}[t]
    \centering
    \includegraphics[width=0.5\textwidth]{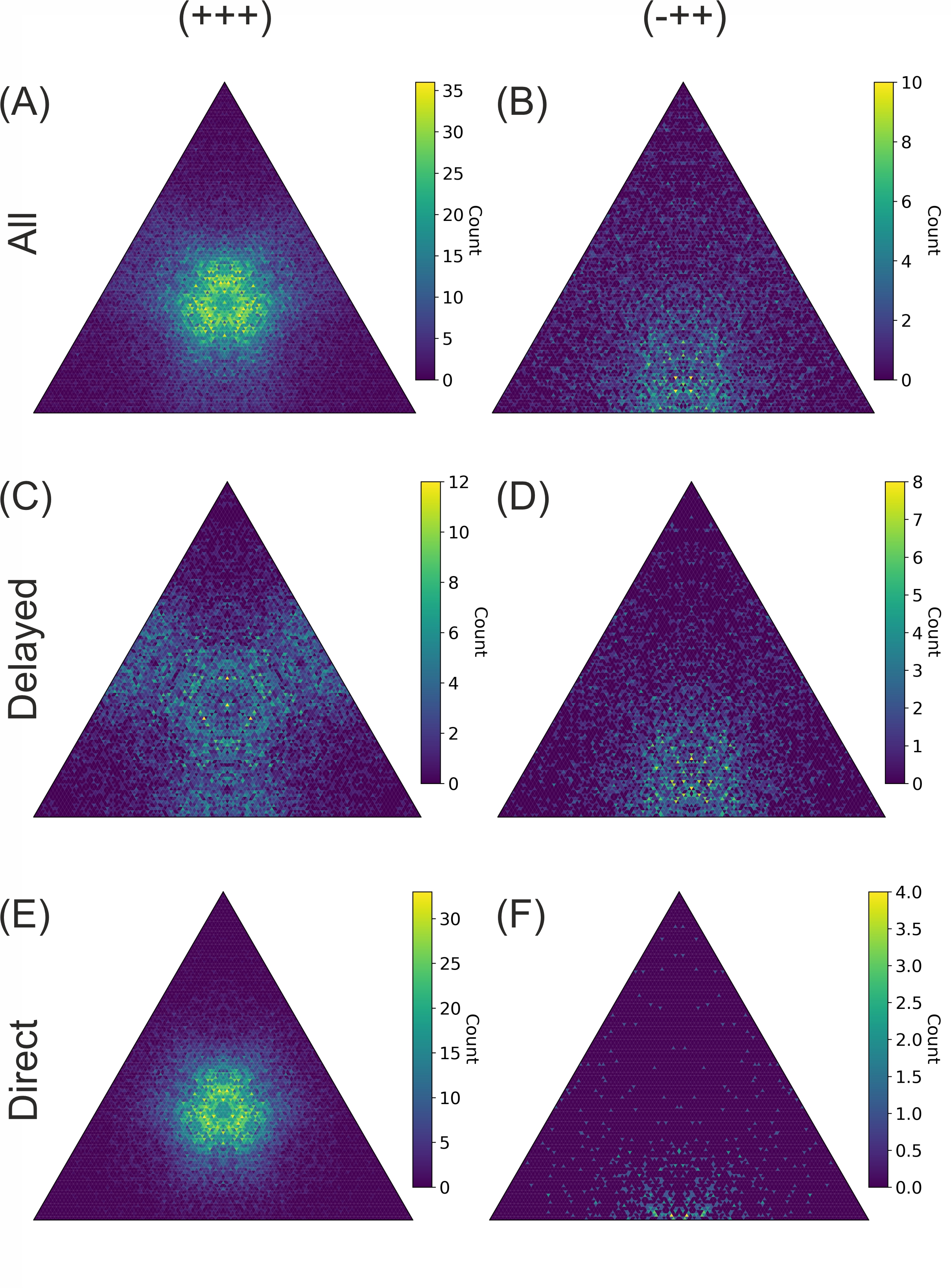}
    \caption{\label{channels}
    Dalitz plots for TI events at $I= 1.3$ ${\rm PW/cm}^2$.  obtained with the ECBB model:(A) and (B) show all events, (C) and (D) show delayed ionization events, and (E) and (F) show direct ionization events. For panels (A), (C) and (E) ternary plots are obtained for all electrons moving in one direction. For panels (B), (D) and (F) ternary plots are obtained for one electron escaping in a direction opposite with respect to the other two electrons (bottom corners of triangle correspond to electrons moving in the same direction). For all panels the classical events  were symmetrized to account for the electrons being indistinguishable.
    }
\end{figure}

While the final state of the escaping electrons is easily read from the distributions in Fig.~\ref{main_feature}, the pathway that leads to this state is not  immediately evident. We attribute features of these distributions to different pathways  the electrons follow to escape by further analyzing the results obtained with the semi-classical ECBB model. In Fig.~\ref{channels}, we present the Dalitz plots  according to the triple ionization pathway the electrons follow to escape.
Panels (A) and (B) (top row) show  all TI events,  panels (E) and (F) (bottom row) show direct TI events, where all electrons escape simultaneously after recollision. Also, panels (C) and (D) (middle row) show  delayed TI events, where two electrons escape immediately after recollision and one escapes with a delay. For the intensities considered  when using the ECBB model, the direct and delayed pathways contribute roughly equally and account for more than 85\% of all events.

\begin{figure*}[t] 
    \centering
    \includegraphics[width=0.9\textwidth]{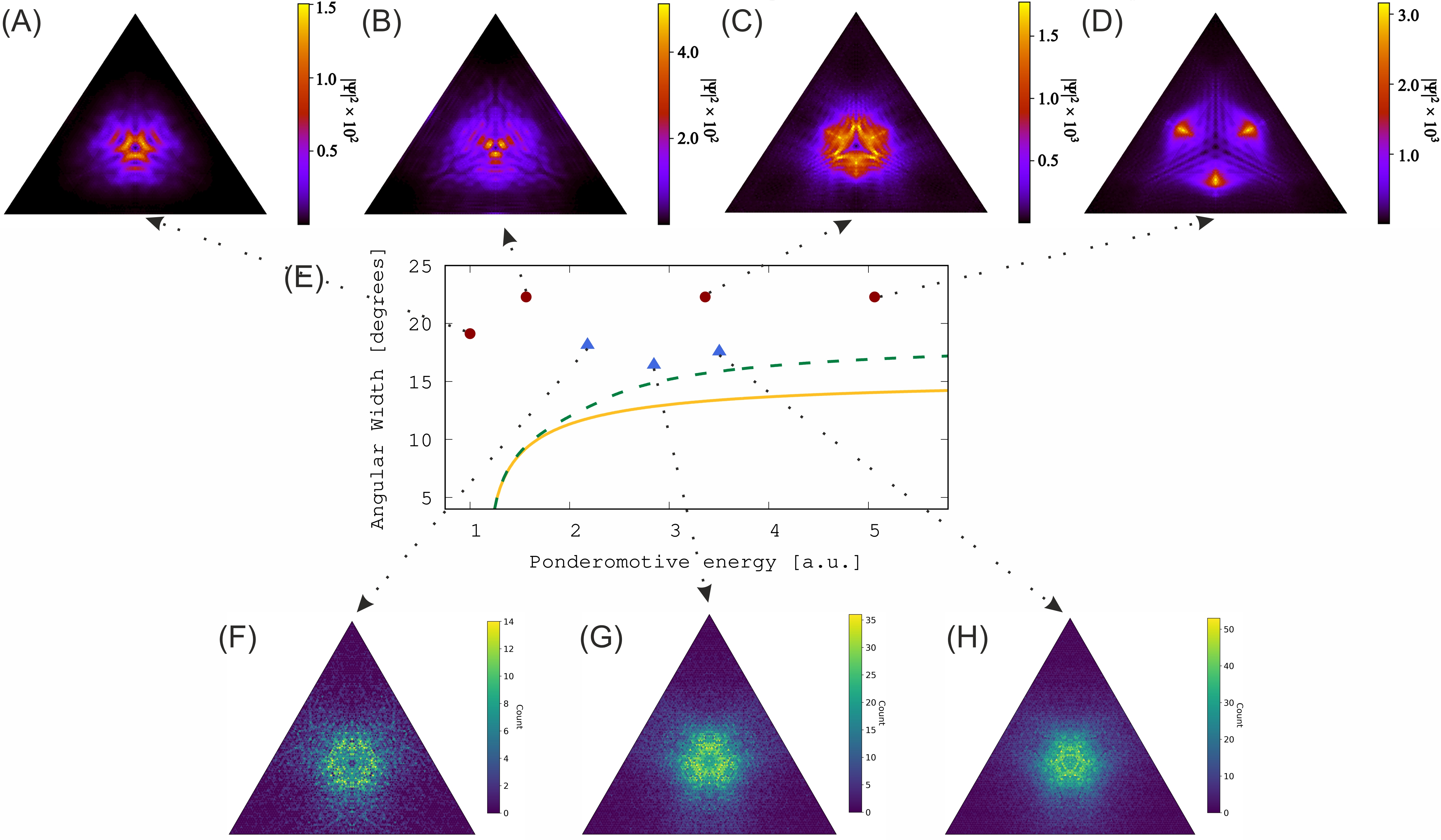}
    \caption{\label{intensity}
    Dalitz plots for data from quantum calculations:
    (A) $I=0.5$ $\rm{PW/cm^2}$, (B) $I=0.75$ ${\rm PW/cm^2}$, (C) $I=1.7$ ${\rm PW/cm^2}$, (D) $I=2.55$ ${\rm PW/cm^2}$.
    Panel (E): Dependence of the angular half-width of three-electron momentum distribution in $(+++)$ octant after direct triple ionization on the value of the ponderomotive energy of the electron gained by the laser field -- the quantum-mechanical model (circles), the classical model (triangles) and simplified analytic models (solid line - $\beta_{3E}^s$, dashed line - $\beta_{3E}$) -- see discussion in the text.
    Dalitz plots for data from ECBB semi-classical calculations: (F) $I=1$ ${\rm PW/cm^2}$, (G)$I=1.3$  ${\rm PW/cm^2}$ and (H) $I=1.6$ ${\rm PW/cm^2}$. }
\end{figure*}

Also,  panels (A), (C) and (E) (left column) show TI events where electrons escape in the same direction along the laser polarization axis, whereas panels (B), (D) and (F) (right column) show TI events where one of the three electrons escapes in a direction opposite to the other two electrons.
Comparing the different rows  in Fig.~\ref{channels}, it is clear that the direct TI events predominantly populate the center of the momentum distribution, while delayed events populate the tails.
In addition, inspection of the right column in Fig.~\ref{channels} reveals that the electron escaping in the opposite direction does so with much lower momentum compared to the other two electrons escaping with similar momenta. What is more, such events correspond mostly to delayed ionization.
 In view of the above associations between features of the Dalitz plots and TI pathways, Fig.~\ref{intensity} reveals that for the ECBB model direct ionization
 becomes more dominant with increasing intensity from 1 PW/cm$^2$ to 1.6 PW/cm$^2$ (see panels (F)--(H) in Fig.~\ref{intensity}). A similar trend is observed for the quantum mechanical calculations with increasing intensity from 0.5 PW/cm$^2$ to 1.7 PW/cm$^2$  (see panels (A)--(D) in Fig.~\ref{intensity}).

Results obtained with the quantum model need some further commenting. In Fig.~\ref{intensity} (panels (A)--(D)) we show momenta distributions for triple ionization events in which all electrons escape in the same direction with respect to the laser polarization axis. First, the model we use, was designed to trace direct ionization by restriction of the space to specified tracks along which electrons may move. These tracks were chosen to coincide with lines along which classical saddle points of potential move when the field intensity is changed (see discussion on saddles in Refs~\cite{sacha2001pathways,eckhardt06} and its application in Refs~\cite{Prauzner-Bechcicki2007,Prauzner-Bechcicki2008,Thiede2018}). Such a choice of track favors direct ionization making it more visible, while keeping overall very good agreement with experiments~\cite{Efimov2018}. Second, the quantum calculations are performed for very short pulses ($n_c=3$) and for single CEP ($\phi=0$). Thus, concerning comparison with semi-classical results,  one may only compare overall trends, since the quantum  momenta distributions are phase-dependent~\cite{Eckhardt2010} due to the short laser-pulses employed. 
Moreover, in the quantum case, disentangling the three-electron momenta distribution into parts that correspond to the direct and delayed paths of ionization, in a similar way as is done for semi-classical results (see Fig.~\ref{channels}), is beyond capabilities offered by our approach. However, the quantum and semi-classical (mainly the ECBB model) Dalitz plots resemble each other. Indeed, they both have  a central ``spot" that does not change its width significantly as the field intensity changes (see Fig.~\ref{intensity}(E)). Thus one can argue that the direct ionization channel indeed may be contained within the quantum Dalitz (+++) plots, and even dominate the distributions there for all studied intensities except the highest one.
For the intensity $I=2.55\; {\rm PW/cm^2}$, panel (D) in Fig.~\ref{intensity}, three separate maxima located at some distance from the center of the triangle are visible, suggesting stronger contribution of delayed ionization.
Resemblance between panels (C) and (H), i.e. between the quantum and the semi-classical Dalitz plots, is striking.

An interesting conclusion can be made about the role of soft-core potentials in strong-field 
 simulations. Soft-core potentials are very useful and often are the first-choice tools for simulations, both in classical and quantum realms. As an example,  we recall the work by Majorosi, Benedict, and Czirj\'ak~ \cite{Majorosi18}, where HHG obtained with a 3D Hydrogen model including the Coulomb potential was successfully reproduced with a single-dimensional soft-core potential model. However, the first choice is not necessarily the finest one. The Heisenberg potential (H model) effectively leads to softening of the Coulomb singularity~\cite{PhysRevA.105.043102,PhysRevA.107.L041101}, therefore in the following discussion the H model will serve as an example of a soft-core semi-classical model.
 At first glance, the general shapes of momentum distributions obtained by both H model and ECBB are similar, and one is tempted to assume that the particular choice of potentials for simulations should not change the qualitative picture of triple ionization. Nevertheless, more careful analysis of  Fig.~\ref{main_feature} reveals that correlated escapes are less prominent in panels (B) and (C), than in panel (A). When these results are compared with the quantum results in Fig.~\ref{main_feature}(D), it is justified to conclude that the ECBB model is better suited for tracing correlations during multi-electron escape. Let us recall that the quantum model considered here is designed in a way that favours  the direct (correlated) escape despite the fact that it uses the soft-core potential. Whether there are better potentials in the case of simplified quantum models remains an open issue.

\subsection{The central ''spot'' in Dalitz plots}

The correspondence of the central ``spot'' in Dalitz (+++) plots mainly with the direct triple ionization channel has been discussed in the previous subsection. That discussion can be  culminated to four important conclusions/conjectures: 
(i) the ``spot'' in semi-classical (+++) Dalitz plots is unambiguously connected with direct triple ionization; 
(ii) semi-classical and quantum (+++) Dalitz plots are quite similar; 
(iii) it is possible to argue for a general correspondence between semi-classical and quantum notions of direct ionization, and
(iv) the size of the ``spot'' is nearly independent for a wide interval of field intensities for both quantum and semi-classical models.
Since the central ``spot'' is formed predominantly by the direct channel, then a simplified classical model for  direct triple ionization is expected to reproduce especially the last property.

In what follows, we construct such a simple classical model for direct triple ionization events and show that this model indeed predicts a nearly constant size of a ``spot'' in Dalitz (+++) plots. Recalling the three-step model, initially one electron tunnel-ionizes close to the field maximum. Then, it propagates in the field and can turn back  and recollide with its parent ion. Upon recollision, the energy gained by the electron in the laser field minus the energy spent to overcome the ionization potential  
is redistributed among all three electrons, i.e. among the tunnel-ionizing and the two bound electrons. In a direct triple ionization event, the electrons are released soon after recollision. Also, we consider that triple ionization takes place following just one recollision, which is supported by our semi-classical calculations \cite{PhysRevA.107.L041101}. In this simplified classical model, we ignore the Coulomb potential and any electron-electron interaction. Hence, as a result of a single recollision in a direct triple ionization event, an electron can gain  a momentum p$_{0}$, due to the energy redistributed by the tunnel-ionizing electron upon recollision. However,  an electron also gains an additional 
momentum $\Delta p$ - a boost from the electric field of the laser pulse, which is equal to  minus the value of the vector potential at the time of ionization, similar to attosecond streaking \cite{Yakovlev2005}. In this simplified model, the momentum shift $\Delta p$ of each electron is the same, thus, in the case when all electrons escape in the same direction,
 for instance (+++), the final momentum distribution is simply shifted from p$_0$.

\begin{figure}[t] 
	 \includegraphics[width=0.4\textwidth]{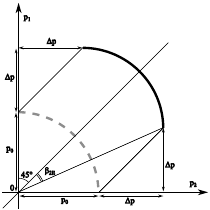}
	 \caption{Schematic illustration of the momentum distribution that electrons gain upon a single recollision in the  laser field. See text for explanation of values depicted here.}
   \label{scheme2E}
\end{figure}

It is easiest to first illustrate the above ideas for double ionization, see Fig.~\ref{scheme2E}.
If the energy of the returning electron  is $U_{r}$ and the ionization potential of the bound electron is $I_{p2}$ (2nd ionization potential), then the total energy that is shared among the two electrons just after recollision is $U_r - I_{p2}$. For a monochromatic laser field, 
the maximum total energy  is  $E_{tm}=3.17U_p  - I_{p2}$.  Also, the additional momentum 
 $\Delta p$ is given by
 $\pm 2\xi \sqrt{U_p}$, where $\xi$ accounts for the recollision time shift with respect to the time when the vector potential has its maximum \cite{Becker2002-ga}. For the case of an electron returning to the ion with the maximum possible energy, i.e. $3.17U_p$,  one obtains approximately  $\xi \approx 0.95$.   Hence, in the momentum space, any direct double ionization event is placed  in the right upper quarter (or left bottom) of the circle $p_1^2+p_2^2=2E_{tm}$, see grey-dashed line in  Fig. \ref{scheme2E}. Results are similar for both quarters so we only focus on the right upper one, see Fig. \ref{scheme2E}.   The radius of the circle is $p_0 = \sqrt{2E_{tm}}$. Following ionization, the laser field  is shifting the electron momentum  by $\Delta p=2\xi \sqrt{U_p}$, giving rise to the black-solid line in Fig. \ref{scheme2E}.   It is clear from the geometry in  Fig. \ref{scheme2E} that the angle between the diagonal and one of the black-solid-line  end points is equal to 
\begin{multline}
\beta_{2E} = \cfrac{\pi}{4} - \arctan{\cfrac{\Delta p}{\Delta p + p_0}} = \\ \cfrac{\pi}{4} - \arctan{\cfrac{2\xi \sqrt{U_p}}{2\xi \sqrt{U_p} + \sqrt{2(3.17U_p - I_{p2})}}}.
\label{a2E}
\end{multline}
 This angle  corresponds to the maximum deviation from the diagonal where both electrons have the same momentum.
Hence,  for double ionization the distribution of direct ionization events will be within the angle $2\beta_{2E}$ from eq. (\ref{a2E}).

Following a similar reasoning, we next obtain the corresponding angle $\beta$  for direct triple ionization events. Direct events  are placed in the $(+++)$ and $(- - -)$ octants of a sphere in momentum space. We illustrate our ideas focusing on the octant $(+++)$. In this case, the angle $\beta_{3E}^s$ is obtained in a similar manner as $\beta_{2E}$ the  difference being that the angle  $\arctan (\Delta p / [\sqrt{2}\Delta p+p_0])$ is not subtracted from $\pi/4$ but from the angle  between the main diagonal of the octant and its bottom plane, which is $\arctan (\sqrt{2}/2)$. Also, the momentum $p_0$ is determined in a similar way as for two electrons, i.e.  $p_0 = \sqrt{2(3.17U_p  - I_{p23})}$, where 
 $I_{p23}$ stands for the sum of the 2nd and 3rd ionization potentials; for  Ne  considered here  $I_{p23}=3.83$ a.u. Thus, we obtain 

\begin{multline}
\beta_{3E}^s = \arctan \left( \cfrac{\sqrt{2}}{2}\right) -  \\ \arctan{\cfrac{1.9 \sqrt{U_p}}{1.9 \sqrt{2U_p} + \sqrt{2(3.17 U_p  - I_{p23})}}}.
\label{a3Es}
\end{multline}

\begin{figure}[t] 
	 \includegraphics[width=0.45\textwidth]{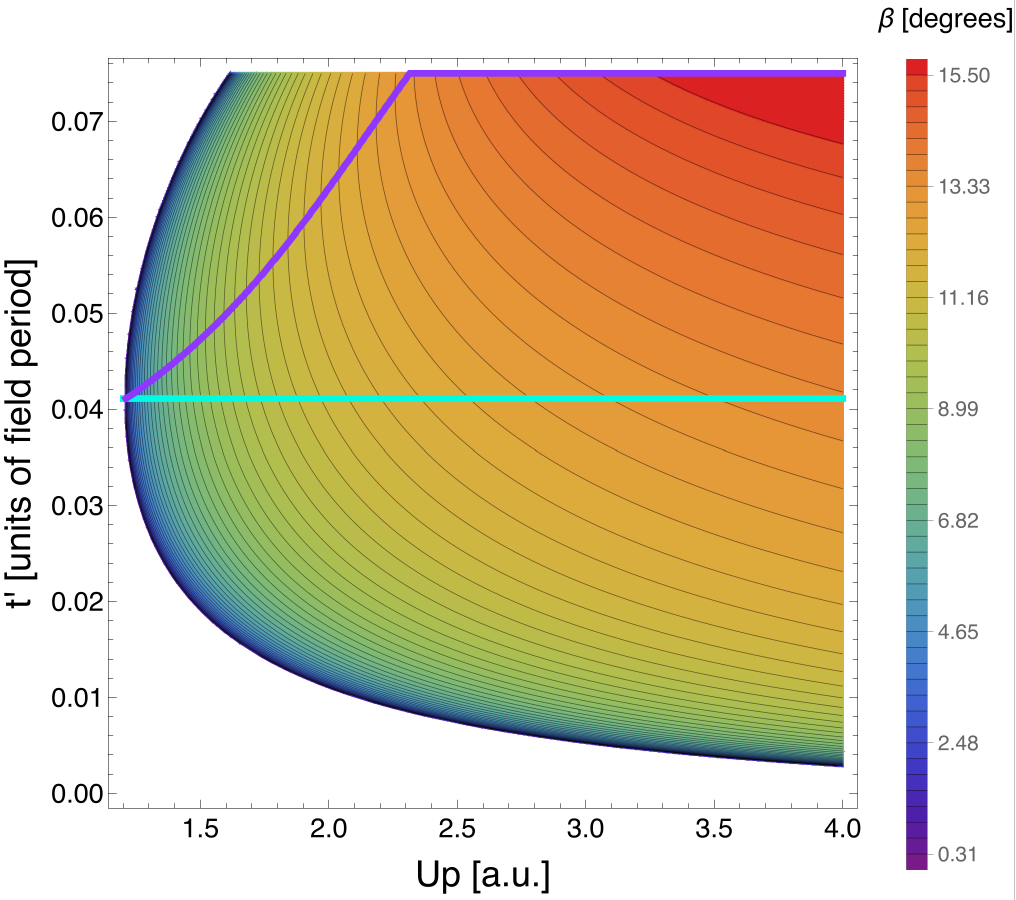}
	 \caption{Angle $\beta_{3E}$ dependence on the time of tunnel-ionization  of the first electron and on ponderomotive energy. The purple curve identifies the value of tunnel-ionization time  that corresponds to the  largest possible $\beta_{3E}$ for each $U_p$. The turquoise curve identifies the angle $\beta_{3E}^s$ that corresponds to constant parameters $\xi=0.95$ and $\kappa=3.17$. }
   \label{contour}
\end{figure}

In Fig.~\ref{intensity}(E) the solid yellow line shows the dependence of the angle $\beta_{3E}^s$ on  $U_p$. Although, the curve does not fit the exact values of the angle extracted from our simulations, it shows the same trend, that is, first the angle grows and then it saturates. Next, we refine the simple model described above in order to better fit the classical results. Until now, we assumed that for all triple ionization events, the tunnel-ionizing electron, upon recollision,  has the maximum possible energy, i.e. $\kappa U_p$, $\kappa = 3.17$. Yet, the largest angle $\beta_{3E}$ does not necessary correspond to the recollision taking place with the largest $\kappa$. Indeed, the factors $\xi$ and $\kappa$ are completely defined by the ponderomotive energy and the time of tunnel-ionization of the recolliding electron. The angle $\beta_{3E}$ is given by

\begin{multline}
\beta_{3E} = \arctan \left( \cfrac{\sqrt{2}}{2}\right) -  \\ \arctan{\cfrac{2\xi(U_p) \sqrt{U_p}}{2\xi(U_p) \sqrt{2U_p} + \sqrt{2(\kappa(U_p) U_p  -  I_{p23})}}}.
\label{a3E}
\end{multline}
In Fig. \ref{contour},  we plot the angle  $\beta_{3E}$ as a function of the time the recolliding electron tunnel-ionizes and as a function of $U_{p}$. We take the time of tunnel-ionization to vary between the time corresponding to the maximum of the laser field (0 time) and 0.075 T forward in time, where T stands for the period of the laser field.  The purple curve corresponds to the angle $\beta_{3E}$ being maximum while the turquoise curve corresponds to the angle $\beta_{3E}$ when $\kappa=3.17$. Using the values obtained in the purple curve for the angle $\beta_{3E}$ as a function of ponderomotive energy, we obtain the   dashed-green line in Fig.~\ref{intensity}(E). This curve passes much closer to values obtained from the semiclassical simulations and still saturates for large values of $U_p$. The remaining discrepancies most probably stem from the  simplified model not accounting for  electron-electron and electron-ion interactions. Hence, this simplified model reproduces  the main features of the angle $\beta_{3E}$ as a function of $U_{p}$. This suggests   that the width of the  momentum distribution  mainly depends on the time of tunnel-ionization (simplified model) and subsequently on the time of recollision and hence does not significantly change with the intensity of the field.

The quantum results reflect the same trend, i.e. nearly constant size of width of the momentum distribution when the intensity increases. Unsurprisingly, the measured values of $\beta$ obtained from quantum simulations do not fit well our simplified classical model. The quantum calculations are performed  using a reduced dimensionality model, therefore one can only expect a qualitative agreement with the simplified classical model.

\section{Conclusions}

We have provided an analysis of signatures of particular triple ionization channels in Dalitz plots of the electron momenta. We have compared data obtained after simulations with a quantum-mechanical restricted-space model with data obtained with semiclassical simulations with the ECBB and Heisenberg models. After a qualitative comparison of the corresponding Dalitz plots, we find  that the results for all numerical models suggest that a central spot in Dalitz plots for electrons propagating all in the same direction correspond to direct ionization. We have rationalized our hypothesis, constructing a simple classical model for direct ionization that predicts how the spot size changes  with increasing intensity of the laser field. The model independence of this imprint of direct ionization manifested on Dalitz plots, suggests that the occurrence of this central spot in Dalitz plots  should also be observed  in experiments.

\begin{acknowledgments}
We gratefully acknowledge Polish high-performance computing infrastructure PLGrid (HPC Centers: ACK Cyfronet AGH) for providing computer facilities and support within computational grant no. PLG/2024/017146. This work was realized under National Science Centre
  (Poland) project Symfonia No. 2016/20/W/ST4/00314.
  JP-B would like to thank Michal Ojczenasz for help in preparing the script for obtaining ternary plots.
\end{acknowledgments}

\bibliographystyle{apsrev4-1}
\bibliography{paper_bib}
\end{document}